\documentstyle[12pt]{article}
\newcommand{\R}{\rm I \mkern -3mu R}
\newcommand{\N}{\rm I \mkern -3muN}
\begin{document}
\thispagestyle{empty}
\begin{flushright}
IFA-FT-393-1994, January
\end{flushright}
\bigskip\bigskip\begin{center}
{\bf \Huge{HAMILTONIAN  MECHANICS IN 1+2 DIMENSIONS}}
\end{center}
\vskip 1.0truecm
\centerline{\bf
D. R. Grigore\footnote{e-mail: grigore@roifa.bitnet}}
\vskip5mm
\centerline{Dept. Theor. Phys., Inst. Atomic Phys.,}
\centerline{Bucharest-M\u agurele, P. O. Box MG 6, ROM\^ANIA}
\vskip 2cm
\bigskip \nopagebreak \begin{abstract}
\noindent
A complete list of all transitive symplectic manifolds
of the Poincar\'e and Galilei group in
1+2 dimensions is given.
\end{abstract}

\newpage\setcounter{page}1

\section{Introduction}

Recently we have given a complete analysis of the projective
unitary irreducible representations of the Poincar\'e and
Galilei groups in 1+2 dimensions [1], [2]. In the context of
constant interest in physics in 1+2 dimensions, we think it
useful to provide the same analysis in the framework of
Hamiltonian mechanics. Namely, we intend to provide a complete
list of all the transitive symplectic actions for the Poincar\'e
and Galilei groups in 1+2 dimensions. The method used is,
esentially, the orbit method of Kostant, Souriau and Kirillov
[3-6]. However, if we want the complete list of these symplectic
actions (not only up to a covering like in the usual formulation
of the orbit method) we need a generalization of this method
appearing in [7].

In Section 2 we present the method following [7]. In Section 3
and 4 we apply the method for the Poincar\'e and the Galilei
groups respectively. The first case is rather standard and
offers no surprises. The second case is on the contrary a very
good "laboratory" for practically all the symplectic techniques
described in full generality in Section 2.

\section{Transitive Symplectic Actions of Lie Groups}

\subsection*{A. Basic Definitions}

Let
$(M_{i},\Omega_{i})~~i=1,2$
be symplectic manifolds. A diffeomorphism
$\phi: M_{1} \rightarrow M_{2}$
is called {\it symplectic} if:
$$
\phi^{*}\Omega_{2} = \Omega_{1}.\eqno(2.1)
$$

Let
$(M,\Omega)$
be a symplectic manifold and $G$ a Lie group (not necessarely
connected) acting on M:
$
G \ni g \mapsto \phi_{g} \in Diff(M).
$

This action is called {\it symplectic} and
$(M,\Omega)$
is called a $G$-{\it symplectic manifold} if:
$$
(\phi_{g})^{*} \Omega = \Omega,~~\forall g \in G.\eqno(2.2)
$$

If
$(M_{i},\Omega_{i})~~i=1,2$
are two $G$-symplectic manifolds, they are called $G$-{\it
symplectomorphic} if there exists a symplectic map
$\phi: M_{1} \rightarrow M_{2}$
which is also $G$-equivariant.

Physically, a $G$-symplectic manifold can be considered as the
phase space of a given Hamiltonian system for which $G$ is the
covariance group. Two such Hamiltonian system are identical from
the physical point of view if the corresponding $G$-symplectic
manifolds are $G$-symplectomorphic. One can also argue that
elementary systems are described by transitive $G$-symplectic
manifolds. The orbit method describes transitive $G$-symplectic
manifolds up to $G$-symplectomorphisms.

\subsection*{B. The Basic Theorem}

We denote by
$Lie(G) \equiv T_{e}(G)$
the Lie algebra of $G$, by
$G \ni g \mapsto Ad_{g} \in End(Lie(G))$
the adjoint action and by
$G \ni g \mapsto Ad_{g}^{*} \in End((Lie(G))^{*})$
the coadjoint action of $G$. A $2$-{\it cocycle} for
$Lie(G)$
is a bilinear antisymmetric map:
$\sigma: Lie(G) \times Lie(G) \rightarrow \R$
verifying the cocycle identity [6]:
$$
\sigma([X_{1},X_{2}],X_{3}) + cyclic~permutations = 0~~\forall
X_{i} \in Lie(G)~(i=1,2,3).\eqno(2.3)
$$

We denote by
$Z^{2}(Lie(G),\R)$
the linear space of al $2$-cocycles. If
$\sigma \in Z^{2}(Lie(G),\R)$
we define:
$$
h_{\sigma} \equiv \{ X \in Lie(G) \vert \sigma(X,Y) = 0, \forall
Y  \in Lie(G)\}\eqno(2.4)
$$
and one finds out that
$h_{\sigma}  \subset Lie(G)$
is a Lie subalgebra. We denote by
$H_{\sigma} \subset G$
the connected Lie subgroup immersed in $G$ and associated to the
Lie subalgebra
$h_{\sigma}$.

There is a natural action of $G$ on
$Z^{2}(Lie(G),\R)$,
namely the coadjoint action given by:
$$
(Ad_{g}^{*} \sigma)(X_{1},X_{2}) = \sigma(Ad_{g^{-1}}(X_{1}),
Ad_{g^{-1}}(X_{2})).\eqno(2.5)
$$

We denote by
$G_{\sigma} \subset G$
the stability subgroup of
$\sigma$
with respect to this action.
$H_{\sigma}$
is a normal subgroup of
$G_{\sigma}$.
An $Ad^{*}$-orbit {\cal O} in
$Z^{2}(Lie(G),\R)$,
is called {\it regular} if for some
$\sigma \in {\cal O}$
(then for all
$\sigma \in {\cal O}$),
the subgroup
$H_{\sigma}$
is closed.

If $K$ is a Lie group and
$N \subset K$
is a normal subgroup we denote by
$K/N$
the factor Lie group. Two subgroups
$Q,Q' \subset K/N$
are called {\it conjugated} if there exists
$k_{0} \in K$
such that:
$$
Q' = \{ k_{0} k k_{0}^{-1} N \vert kN \in Q\}.\eqno(2.6)
$$

Now we can formulate the main theorem [7]:

{\bf Theorem} Take one representative
$\sigma$
from every regular orbit in
$Z^{2}(Lie(G),\R)$.
Let
${\cal H}_{\sigma}$
be the set of discrete subgroups of
$G_{\sigma}/H_{\sigma}$
and
${\cal C}_{\sigma}$
the set of conjugacy classes in
${\cal H}_{\sigma}$.
Let
$\bar{H} \in {\cal H}_{\sigma}$
be a representative of a given conjugacy class
$[\bar{H}] \in {\cal C}_{\sigma}$
and:
$$
H \equiv \{ h \in G_{\sigma} \vert h H_{\sigma} \in [\bar{H}] \}.\eqno(2.7)
$$
Then
$H \subset G$
is a closed subgroup and
$G/H$
is a $G$-symplectic manifold with the symplectic form
$\Omega^{\sigma}$
uniquely determined by:
$$
\sigma = (\pi^{*} \Omega^{\sigma})_{e}.\eqno(2.8)
$$
(here
$\pi:G \rightarrow G/H$
is the canonical submersion.)

Every $G$-symplectic manifold of $G$ is $G$-symplectomorphic
with a manifold of the form
$(G/H,\Omega^{\sigma})$
described above. Moreover, to different couples
$(\sigma,[\bar{H}]) \not= (\sigma',[\bar{H}'])$
correspond $G$-symplectic manifolds which are not
$G$-symplectomorphic.
\vskip 0.5truecm

This theorem is quite general and affords a complete
classification in a very constructive way. Loosely speaking, the
regular orbits classify the $G$-symplectic manifolds, up to a
covering; the various coverings are classified by some classes
of discrete subgroups in
$G_{\sigma}/H_{\sigma}$.

In applications, we will first determine for every regular
orbit, the {\it maximal} symplectic manifold corresponding to
$H = H_{\sigma}$
in the construction above. Then the various symplectic manifolds
covered by this maximal manifold will be determined by factorizing
$(G/H_{\sigma},\Omega^{\sigma})$
to the (symplectic) action of some suitable chosen discrete
subgroup of $G$.

\subsection*{C. A Particular Case}

A $2$-{\it coboundary} is an element of
$Z^{2}(Lie(G),\R)$
of the form:
$$
\sigma(X_{1},X_{2}) = - <\eta,[X_{1},X_{2}]>,~~\forall
X_{1},X_{2} \in Lie(G).\eqno(2.9)
$$

Here $< , >$ is the duality form between
$Lie(G)$
and
$(Lie(G))^{*}$
and
$\eta \in (Lie(G))^{*}$
is arbitrary. The linear space of all $2$-coboundaries is
denoted by
$B^{2}(Lie(G),\R)$.
We will also need the {\it second cohomology group}
with real coefficients
$$
H^{2}(Lie(G),\R) \equiv Z^{2}(Lie(G),\R)/B^{2}(Lie(G),\R).
$$

Finally, an $1$-{\it cocycle} for
$Lie(G)$
is any element
$\eta \in (Lie(G))^{*}$
verifying:
$$
<\eta,[X_{1},X_{2}]>~= 0,~~\forall X_{1},X_{2} \in Lie(G).\eqno(2.10)
$$

We denote by
$Z^{1}(Lie(G),\R)$
the linear space of all $1$-cocycles and by
$H^{1}(Lie(G),\R) \equiv Z^{1}(Lie(G),\R)$
the {\it first cohomology group} with real coefficients.

For every
$X \in Lie(G)$,
let
$X_{\cal O}$
be the associated vector field on {\cal O}:
$$
X_{\cal O} \equiv {d \over ds} Ad_{exp(-sX)}^{*}\vert_{s=0}.\eqno(2.11)
$$

Then we have:

{\bf Corollary:} Let
${\cal O} \subset (Lie(G))^{*}$
a coadjoint orbit. Then {\cal O} becomes a symplectic manifold
with respect to the Kostant-Souriau-Kirillov symplectic form
$\Omega^{KSK}$
which is uniquely determined by:
$$
\Omega^{KSK}_{\eta}(X_{\cal O},Y_{\cal O}) = -<\eta,[X,Y]>,\eqno(2.12)
$$
($\forall \eta \in {\cal O}, \forall X,Y \in Lie(G)$).

Two different coadjoint orbits are not $G$-symplectomorphic.

Suppose that
$H^{i}(Lie(G),\R) = 0~(i=1,2)$.
If the stability subgroup
$G_{\eta}~(\eta \in {\cal O}$
arbitrary) is connected, then every transitive $G$-symplectic
manifold is $G$-symplectomorphic with a coadjoint orbit. If
$G_{\eta}$
is not connected, then the coadjoint orbits are the maximal
symplectic manifolds and the various symplectic manifolds
covered by {\cal O} are classified by the conjugacy classes in
$G_{\eta}/(G_{\eta})^{0}$.
\vskip 0.5truecm

In applications one can use the factorization method outlined at
the end of Subsection 2B.
Let us remark that if
$H^{i}(Lie(G),\R) \not= 0$
for
$i=1$
or
$i=2$
then the list of all transitive $G$-symplectic manifolds is not
exhausted by the construction outlined above and based on
coadjoint orbits.

\subsection*{D. Extended Coadjoint Orbits}

We close this section with another interesting construction. We
have seen in Subsection 2B that if
$H^{i}(Lie(G),\R) = 0~(i=1,2)$
then one can conveniently describe transitive $G$-symplectic
manifolds as coadjoint orbits of $G$ (or their factorization).
One may wonder if something analogous works in the general case.
The answer is positive and the construction works as follows [4]-[5].

Let $c$ be a $2$-cocycle for the Lie group $G$, i.e. a map
$c:G \times G \rightarrow \R$
verifying:
$$
c(g_{1},g_{2}) + c(g_{1}g_{2},g_{3}) = c(g_{2},g_{3}) +
c(g_{1},g_{2}g_{3})~(\forall g_{1},g_{2},g_{3} \in G),\eqno(2.13)
$$
$$
c(e,g) = c(g,e) = 0~(\forall g \in G).\eqno(2.14)
$$

We construct the {\it central extension}
$G^{c}$
of $G$ which is set-theoretically
$G^{c} = G \times \R$
with the composition law:
$$
(g;\zeta)\cdot(g';\zeta') = (gg';\zeta+\zeta'+c(g,g')).\eqno(2.15)
$$

Then if $c$ is smooth,
$G^{c}$
is also a Lie group. We identify
$Lie(G^{c}) \simeq Lie(G) + \R$
and
$(Lie(G^{c}))^{*} \simeq (Lie(G))^{*} + \R$
in a natural way.

Then one can show that the coadjoint action of
$G^{c}$
has the form:
$$
Ad^{*}_{g;\zeta}(\eta;\rho) =
(Ad^{*}_{g}(\eta)+\rho\alpha_{g^{-1}};\rho) .\eqno(2.16)
$$

Here
$\alpha_{g} \in (Lie(G))^{*}$
is given by:
$$
<\alpha_{g},X> = {d\over ds}\left[ c\left( g,e^{sX}\right)-
c\left( e^{sAd_{g}(X)},g\right)\right]\vert_{s=0}.\eqno(2.17)
$$

One notices that the orbits of the action (2.16) are of the form
$({\cal O};\rho)$
where {\cal O} are orbits in
$(Lie(G))^{*}$
relative to a modified coadjoint action. In particular, we
consider the case
$\rho = 1$
and obtain the modified coadjoint action:
$$
Ad^{c}_{g}(\eta) \equiv Ad^{*}_{g}(\eta) + \alpha_{g^{-1}}.\eqno(2.18)
$$

Because
$Ad^{*}_{g}$
is modified only by an
$\eta$-
independent translation, it is clear that
$Ad^{c}_{g}$
will remain a symplectic transformation with respect to
$\Omega^{KSK}$.
It follows that in this way we obtain new coadjoint orbits as
transitive $G$-symplectic manifolds.

One can prove that two construction of this type, based on the
$2$-cocycles
$c_{1}$
and
$c_{2}$
respectively, give the same result (up to a
$G$-symplectomorphism) {\it iff}
$c_{1}$
and
$c_{2}$
are cohomologous, i.e:
$$
c_{1}(g,g') - c_{2}(g,g') = d(g) - d(gg') + d(g')\eqno(2.19)
$$
for some smooth
$d:G \rightarrow \R$.

In conclusion on can obtain new $G$-symplectic manifolds (beside
the usual coadjoint orbits), by classifying all $2$-cocycles of
$G$, up to the equivalence relation (2.19), selecting a
representative from every cohomology class and working with the
central extension
$G^{c}$.

There is no guarantee that we will obtain all the transitive
$G$-symplectic manifolds in this way although this happens, for
instance, for the Galilei group in 1+3 dimensions. In fact the
Galilei group in 1+2 dimensions provide an example for which we
do not obtain all the transitive $G$-symplectic manifolds in
this way. The list can be however completed by some simple
tricks performed on the coadjoint orbits.

\section{Transitive Symplectic Actions for the Poincar\'e Group
in 1+2 Dimensions}

We denote by $M$ the 1+2-dimensional Minkowski space i.e.
$\R^{3}$
with coordinates
$(x^{0},x^{1},x^{2})$
and with the Minkowski bilinear form:
$$
\{x,y\} \equiv x^{0} y^{0} - x^{1} y^{1} - x^{2} y^{2}.\eqno(3.1)
$$

We will also need the Minkowski norm:
$\Vert x\Vert \equiv \{x,x\}$

The Lorentz group is:
$$
{\cal L} \equiv \{ \Lambda \in End(M) \vert \{Lx,Ly\} =
\{x,y\}, \forall x,y \in M \}
$$
considered as group with respect to operator multiplication.

The proper orthochronous Lorentz group is:
${\cal L}^{\uparrow}_{+} \subset {\cal L}$:
$$
{\cal L}^{\uparrow}_{+} \equiv \{ \Lambda \in {\cal L} \vert
det(L)=1,~L_{00}>0\}
$$

The proper orthochronous Poincar\'e group is a semi-direct product:
set theoretically
${\cal P}^{\uparrow}_{+}$
is formed from couples
$(L,a)$
with
$L \in {\cal L}^{\uparrow}_{+}$
and
$a \in M$
and the composition law is:
$$
(L,a)\cdot (L',a') = (LL',a+La').\eqno(3.2)
$$

It is well known that:
$
H^{i}(Lie({\cal P}^{\uparrow}_{+}),\R) = 0~~(i=1,2)
$
(see e.g. [6]) so we can apply the corollary from Subsection 2C.

One can identify
$(Lie({\cal P}^{\uparrow}_{+}))^{*} \simeq \wedge^{2}M+M$
(see [6]). One can naturally extend to this space the action of
${\cal P}^{\uparrow}_{+}$,
the Minkowski bilinear form
$\{ , \}$
and the Minkowski norm
$\Vert\cdot\Vert$.
The coadjoint action is then given by:
$$
Ad^{*}_{L,a}(\Gamma,P) = (L\Gamma+a\wedge LP,LP).\eqno(3.3)
$$

One can easily compute the coadjoint orbits of
${\cal P}^{\uparrow}_{+}$.
If
$e_{0}, e_{1}, e_{2}$
is the canonical base in $M$, then they are:

(a)
$
M_{m,s}^{\epsilon} \equiv \{(\Gamma,P) \vert \Vert P\Vert^{2} =
m^{2},~sign(P_{0}) = \epsilon,~\Gamma \wedge P = \epsilon ms
e_{0} \wedge e_{1} \wedge e_{2} \}~~(m \in \R_{+},~s \in \R,~
\epsilon=\pm)
$

(b)
$
M^{\epsilon}_{s} \equiv \{(\Gamma,P) \vert \Vert P\Vert^{2} =
0,~sign(P_{0}) = \epsilon,~\Gamma \wedge P = \epsilon s
e_{0} \wedge e_{1} \wedge e_{2} \}~~(s \in \R,~\epsilon=\pm)
$

(c)
$
M_{m,s} \equiv \{(\Gamma,P) \vert \Vert P\Vert^{2} =
-m^{2},~\Gamma \wedge P = \epsilon ms e_{0} \wedge e_{1} \wedge e_{2} \}
{}~~(m \in \R_{+},~s \in \R)
$

(d1)
$
\widetilde{M}^{\epsilon}_{m} \equiv \{(\Gamma,0)\vert \Vert
*\Gamma\Vert ^{2} = m^{2},~sign((*\Gamma)_{00}) = \epsilon \}~~
(m \in \R_{+},~\epsilon=\pm)
$

(d2)
$
M^{\epsilon} \equiv \{(\Gamma,0)\vert \Vert *\Gamma\Vert ^{2} = 0,~
sign((*\Gamma)_{00}) = \epsilon \}~~(\epsilon=\pm)
$

(d3)
$
M_{m} \equiv \{(\Gamma,0)\vert \Vert *\Gamma\Vert ^{2} = -m^{2} \}~~
(m \in \R_{+})
$

(here $*$ is the Hodge operator.)

Computing the stability subgroups for a given reference point
from every orbit we obtain only connected Lie subgroups.
Applying the corollary from Subsection 2C it follows that
(a)-(d) is the complete list of the transitive
${\cal P}^{\uparrow}_{+}$-
symplectic manifolds.

{\bf Remark 1:} A different realization of
$M^{\epsilon}_{m,s}$
also appeared in [8]. It is interesting to establish the
connection between these two realizations. The idea is to identify
$$
\wedge^{2}M \ni \Gamma \leftrightarrow J = *\Gamma \in M.\eqno(3.4)
$$

Then the action (3.3) becomes:
$$
Ad^{*}_{L,a}(J,P) = (LJ+a \times LP,LP)\eqno(3.5)
$$
where, for any
$a,b \in M$
we define
$a \times b \in M$
according to:
$$
(a \times b)^{\rho} \equiv \varepsilon^{\rho\mu\nu} a_{\mu}
b_{\nu}.\eqno(3.6)
$$

Then the manifolds (a)-(d) above become subsets of points
$(J,P) \in M \times M$:

(a)
$
M_{m,s}^{\epsilon} \equiv \{(J,P) \vert \Vert P\Vert^{2} =
m^{2},~sign(P_{0}) = \epsilon,~\{J,P\} = \epsilon ms\}~~
(m \in \R_{+},~s \in \R,~
\epsilon=\pm)
$

(b)
$
M^{\epsilon}_{s} \equiv \{(J,P) \vert \Vert P\Vert^{2} =
0,~sign(P_{0}) = \epsilon,~\{J,P\} = \epsilon s\}~~
(s \in \R,~\epsilon=\pm)
$

(c)
$
M_{m,s} \equiv \{(J,P) \vert \Vert P\Vert^{2} =
-m^{2},~\{J,P\} =  ms \}
{}~~(m \in \R_{+},~s \in \R)
$

(d1)
$
\widetilde{M}^{\epsilon}_{m} \equiv \{(J,0)\vert \Vert
J\Vert ^{2} = m^{2},~sign(J_{00}) = \epsilon \}~~
(m \in \R_{+},~\epsilon=\pm)
$

(d2)
$
M^{\epsilon} \equiv \{(J,0)\vert \Vert J\Vert ^{2} = 0,~
sign(J_{00}) = \epsilon \}~~(\epsilon=\pm)
$

(d3)
$
M_{m} \equiv \{(J,0)\vert \Vert J\Vert ^{2} = -m^{2} \}~~
(m \in \R_{+})
$

{\bf Remark 2:} One can investigate now the notion of
localisability for the systems described above following the
lines of [9]. It is not hard to establish that only the system
corresponding to
$M^{\epsilon}_{m,0}$
can be localisable, namely on the Euclidean space
$\R^{2}$.

{\bf Remark 3:} In all cases the identity map is a {\it bona
fide} momentum map.

{\bf Remark 4:} If we compare the actions above with the list of
projective unitary irreducible representations of the same
group [1] it is appearent that there are representations which
do not have a classical analogue.

\section{The Transitive Symplectic Actions for the
Galilei Group in 1+2 Dimensions}

\subsection*{A. Notations}

We define directly the proper ortochronous Galilei group
in $1+2$ dimensions
${\cal G}^{\uparrow}_{+}$
as the group of
$4 \times 4$
real matrices of the form:
$$
(R,{\bf v},\tau,{\bf a}) \equiv
\left( \matrix{ R & {\bf v} & {\bf a} \cr 0 & 1 & \tau \cr 0 & 0
& 1} \right)\eqno(4.1)
$$
where
$\tau \in \R$,
$R \in SO(2)$
is a
$2 \times 2$
real orthogonal matrix and the vectors
${\bf v}, {\bf a} \in \R^{2}$
are considered as column matrices.

As for any matrix group, we identify the Lie algebra
$
Lie({\cal G}^{\uparrow}_{+})
$
with the linear space of
$
4 \times 4
$
real matrices of the form:
$$
(\alpha,{\bf u},t,{\bf x}) \equiv \left( \matrix{\alpha A & {\bf
u} & {\bf x} \cr 0 & 0 & t \cr 0 & 0 & 0}\right).
\eqno(4.2)
$$

Here
$
{\bf u}, {\bf x} \in \R^{2},~t, \alpha \in \R,~
A \equiv \left( \matrix{ 0 & 1 \cr -1 & 0}\right),
$
and the exponential map is the usual matrix exponential. One can
easily obtains the Lie bracket as:
$$
[(\alpha_{1},{\bf u}_{1},t_{1},{\bf x}_{1}),
(\alpha_{2},{\bf u}_{2},t_{2},{\bf x}_{2})] =
(0,A(\alpha_{1} {\bf u}_{2}-\alpha_{2} {\bf u}_{1}),0,
A(\alpha_{1} {\bf x}_{2}-\alpha_{2} {\bf x}_{1})+
t_{2} {\bf u}_{1}-t_{1} {\bf u}_{2}).
\eqno(4.3)
$$

We have established [2] that
$H^{2}(Lie({\cal G}^{\uparrow}_{+}),\R) \not= 0
$
(in fact it is a three-dimensional real space) so we will have
to apply directly the theorem from Subsection 2B.

First we choose a convenient representation for an arbitrary
element from
$Z^{2}(Lie({\cal G}^{\uparrow}_{+}),\R)$.
{}From [2] it follows that a generic element is of the form:
$[m,F,S,{\bf G},{\bf P}]~(m, F, S \in \R,~{\bf G}, {\bf P} \in \R^{2}$)
given by the following formula:
$$
[m,F,S,{\bf G},{\bf P}] = m \xi_{0} + F \xi_{1} + S \xi_{2} + [{\bf
G},{\bf P}]\eqno(4.4)$$
where
$\xi_{0}, \xi_{1}, \xi_{2}$
are non-trivial cocycles (i.e. they are not coboundaries) and they
have the following expressions:
$$
\xi_{0}((\alpha_{1},{\bf u}_{1},t_{1},{\bf x}_{1}),
(\alpha_{2},{\bf u}_{2},t_{2},{\bf x}_{2})) = {\bf x}_{1}\cdot
{\bf u}_{2} - {\bf x}_{1}\cdot {\bf u}_{2}\eqno(4.5)
$$
$$
\xi_{1}((\alpha_{1},{\bf u}_{1},t_{1},{\bf x}_{1}),
(\alpha_{2},{\bf u}_{2},t_{2},{\bf x}_{2})) =
<{\bf u}_{1},{\bf u}_{2}>\eqno(4.6)
$$
$$
\xi_{2}((\alpha_{1},{\bf u}_{1},t_{1},{\bf x}_{1}),
(\alpha_{2},{\bf u}_{2},t_{2},{\bf x}_{2})) =
\alpha_{1} t_{2} - \alpha_{2} t_{1}\eqno(4.7)
$$
and
$[{\bf G},{\bf P}]$
is a coboundary of the form:
$$
[{\bf G},{\bf P}] ((\alpha_{1},{\bf u}_{1},t_{1},{\bf x}_{1}),
(\alpha_{2},{\bf u}_{2},t_{2},{\bf x}_{2})) =
$$
$$
<{\bf P},\alpha_{1}{\bf x}_{2}-\alpha_{2}{\bf x}_{1}> +
{\bf P}\cdot(t_{2}{\bf u}_{1}-t_{1}{\bf u}_{2}) -
<{\bf G},\alpha_{1}{\bf u}_{2}-\alpha_{2}{\bf u}_{1}>.\eqno(4.8)
$$

We have denoted the usual scalar product in
$\R^{2}$
by
${\bf x}\cdot {\bf y}$
and
$<~,~>$
is the symplectic form on
$\R^{2}$:
$$
<{\bf x},{\bf y}> \equiv {\bf x}\cdot A{\bf y}.\eqno(4.9)
$$

\subsection*{B. Coadjoint Orbits in
$Z^{2}(Lie({\cal G}^{\uparrow}_{+}),\R)$}

We need the corresponding coadjoint action. First, we compute
from (4.1) and (4.2) the adjoint action:
$$
Ad_{R,{\bf v},\tau,{\bf a}}(\alpha,{\bf u},t,{\bf x}) =
(R,{\bf v},\tau,{\bf a})(\alpha,{\bf u},t,{\bf x})
(R,{\bf v},\tau,{\bf a})^{-1} =
$$
$$
(\alpha,R{\bf u}-\alpha A{\bf v},t,R{\bf x}+t{\bf v}+\alpha
A(\tau {\bf v}-{\bf a})-\tau R{\bf u}).\eqno(4.10)
$$

Then, applying (2.4) we get the desired coadjoint action:
$$
Ad^{*}_{R,{\bf v},\tau,{\bf a}}[m,F,S,{\bf G},{\bf P}] =
[m,F,S,R({\bf G}+m{\bf a})-\tau R({\bf P}+m{\bf v})-FA{\bf v},
R{\bf P}+m{\bf v}].\eqno(4.11)
$$

It is clear that the structure of the coadjoint orbits will
depend on $F$. In particular we have two cases
$F = 0$
and
$F \not= 0$.

(I) $\underline{F = 0}$

In this case the coadjoint orbits are:

(a)
$
{\cal O}^{1}_{m,s} \equiv \{[m,0,S,{\bf G},{\bf P}] \vert~
{\bf G},{\bf P} \in \R^{2}\}~~m \in \R^{*},~S \in \R
$

(b)
$
{\cal O}^{2}_{s,k,\lambda} \equiv \{[0,0,S,{\bf G},{\bf P}] \vert~
{\bf P}^{2} = k^{2},~{\bf G}\wedge {\bf P}=\lambda k
{\bf e}_{1}\wedge {\bf e}_{2} \}~~k \in \R_{+},~S,\lambda \in \R
$

(c)
$
{\cal O}^{3}_{S,k} \equiv \{[0,0,S,{\bf G},{\bf 0}] \vert~
{\bf G}^{2} = k^{2} \}~~k \in \R_{+}\cup \{0\}
$

(II) $\underline{F \in \R^{*}}$

(a)
$
{\cal O}^{4}_{m,s} \equiv \{[m,F,S,{\bf G},{\bf P}] \vert~
{\bf G},{\bf P} \in \R^{2}\}~~m \in \R^{*},~S \in \R
$

(b)
$
{\cal O}^{5}_{m,s} \equiv \{[0,F,S,{\bf G},{\bf P}] \vert~
{\bf G} \in \R^{2}, {\bf P}^{2}=k^{2}\}~~k \in \R_{+}\cup\{0\},~S \in \R
$

Above we have denoted with
${\bf e}_{1}$
and
${\bf e}_{2}$
the natural basis in
$\R^{2}$.

\subsection*{C. Computation of the Transitive Symplectic Actions}

As indicated in the statement of the theorem from Subsection 2B,
we need to provide a list of subgroups
$H \subset {\cal G}^{\uparrow}_{+}$
such that
${\cal G}^{\uparrow}_{+}/H$
is a symplectic manifold with the symplectic form given by
(2.8). Of course, this is a very implicit way to exhibit the
symplectic transitive actions of
${\cal G}^{\uparrow}_{+}$.

As suggested in Subsection 3B, we divide the study in two cases.
One has to take some reference point
$\sigma$
on every orbit
${\cal O}^{i}~(i=1,...,5)$
and thereafter to compute
$G_{\sigma}$
and
$H_{\sigma}$
and the discrete subgroups of
$G_{\sigma}/H_{\sigma}$.
The computations are elementary and we provide only the final
results. We point out that in all the cases the action of
$G_{\sigma}$
on
$G_{\sigma}/H_{\sigma}$
is trivial, so
${\cal C}_{\sigma} = {\cal H}_{\sigma}$
(in the notations of the theorem from 2B).

(I) \underline {F = 0}

(a)
$\underline{\sigma = [m,0,S,{\bf 0},{\bf 0}]}$

One finds two subcases:

(a1)
$\underline{S = 0}$
$$
H_{\sigma} = \{(R,{\bf 0},\tau,{\bf 0})\vert R \in SO(2),~\tau
\in \R\}
$$

(a2)
$\underline{S \not= 0}$
$$
H_{\sigma} = \{({\bf 1},{\bf 0},0,{\bf 0})\}
$$

In both cases we have
$$
G_{\sigma} = \{(R,{\bf 0},\tau,{\bf 0})\vert R \in SO(2),~\tau
\in \R\}
$$

So, in the case (a1),
$G_{\sigma}/H_{\sigma}$
is trivial and in the case (a2)
$G_{\sigma}/H_{\sigma} \simeq G_{\sigma} \simeq SO(2) \times \R$.

(b)
$\underline{\sigma = [0,0,S,\lambda{\bf e}_{1},k{\bf e}_{2}]}$

One finds:
$$
H_{\sigma} = \{({\bf 1},{\bf v},0,{\bf a})\vert v_{2} = 0,~
a_{1} = 0\}
$$
$$
G_{\sigma} = \{({\bf 1},{\bf v},0,{\bf a})\vert {\bf v},{\bf a}
\in \R^{2}\}
$$
$$
G_{\sigma}/H_{\sigma} \simeq \{({\bf 1},{\bf v},0,{\bf a})\vert
v_{1} = 0,~a_{2} = 0\} \simeq \R \times \R
$$

(c)
$\underline{\sigma = [0,0,S,k{\bf e}_{2},{\bf 0}]}$

Again we have two subcases:

(c1)
$\underline{k \in \R_{+}}$
$$
H_{\sigma} = \left\{({\bf 1},{\bf v},\tau,{\bf a})\vert
v_{1} = -{S\over k},~\tau,v_{2} \in \R,~{\bf a} \in \R^{2}\right\}
$$
$$
G_{\sigma} = \{({\bf 1},{\bf v},\tau,{\bf a})\vert
\tau \in \R,~{\bf v},{\bf a} \in \R^{2}\}
$$
$$
G_{\sigma}/H_{\sigma} \simeq \{({\bf 1},{\bf v},0,{\bf 0})\vert
v_{2} = 0\} \simeq \R
$$

(c2)
$\underline{k=0}$

If
$S \not= 0$,
we have:
$$
H_{\sigma} = \{({\bf 1},{\bf v},0,{\bf a})\vert
{\bf v},{\bf a} \in \R^{2}\}
$$
$$
G_{\sigma} = {\cal G}^{\uparrow}_{+}
$$
$$
G_{\sigma}/H_{\sigma} \simeq \{(R,{\bf 0},\tau,{\bf 0})\vert
R \in SO(2), ~\tau \in \R\} \simeq SO(2) \times \R.
$$

If
$S = 0$,
we have
$G_{\sigma} = H_{\sigma} = {\cal G}^{\uparrow}_{+}$
so this case is trivial.

(II)
$\underline{F \in \R^{*}}$

(a)
$\underline{\sigma = [m,F,S,{\bf 0},{\bf 0}]}$

We obtain the same subgroups
$G_{\sigma}$
and
$H_{\sigma}$
as in the case
$F = 0$.

(b)
$\underline{\sigma = [0,F,S,{\bf 0},k{\bf e}_{2}]}$

We have two subcases:

(b1)
$\underline{k \in \R_{+}}$
$$
H_{\sigma} = \left\{\left({\bf 1},v{\bf e}_{1},{Fv\over k},
\left( {SFv\over k^{2}}+{Fv^{2}\over 2k^{2}}\right)
{\bf e}_{1}+a{\bf e}_{2}\right)\vert v,a \in \R\right\}
$$
$$
G_{\sigma} = \left\{\left({\bf 1},v{\bf e}_{1},{Fv\over k},{\bf a}\right)
\vert v \in \R,~{\bf a} \in \R^{2}\right\}
$$
$$
G_{\sigma}/H_{\sigma} = \{({\bf 1},{\bf 0},0,b{\bf e}_{1})\vert
b \in \R\} \simeq\R
$$

(b2)
$\underline{k = 0}$

There are two possibilities:

(b21)
$\underline{S \not= 0}$
$$
H_{\sigma} = \{({\bf 1},{\bf 0},0,{\bf a})\vert {\bf a} \in \R^{2}\}
$$

(b21)
$\underline{S = 0}$
$$
H_{\sigma} = \{(R,{\bf 0},\tau,{\bf a})\vert R \in SO(2), ~\tau
\in \R,~{\bf a} \in \R^{2}\}
$$

Regardless of the value of $S$ we have:
$$
G_{\sigma} = \{(R,{\bf 0},\tau,{\bf a})\vert R \in SO(2), ~\tau
\in \R,~{\bf a} \in \R^{2}\}
$$

So we have for the first possibility
$$
G_{\sigma}/H_{\sigma} = \{(R,{\bf 0},\tau,{\bf 0})\vert R \in SO(2), ~\tau
\in \R\} \simeq SO(2) \times \R
$$
and for the second possibility the factor group is trivial.

As regards the discrete subgroups of
$G_{\sigma}/H_{\sigma}$
we have only three non-trivial possibilities:
$\R,\R \times \R$
and
$SO(2) \times \R$
as it is appearent from the list above. It is well known that
the discrete subgroups are in these cases
$\bar{H}_{\gamma} \equiv \gamma {\bf Z},
\bar{H}_{\gamma_{1},\gamma_{2}} \equiv \gamma_{1} {\bf Z} \times
\gamma_{2} {\bf Z}$
and
$\bar{H}_{r,\gamma} \equiv Z_{r} \times \gamma {\bf Z}$
respectively. Here
$\gamma, \gamma_{1}, \gamma_{2} \in \R_{+} \cup\{ 0\}$
and
$Z_{r} \in SO(2)$
is the cyclic group of order
$r \in \N^{*}: Z_{r} \equiv \{ R(2\pi k/r) \vert k=0,...,r-1\}$
(as usual
$R(\phi) = e^{\phi A}$
is the rotation of angle
$\phi$).

Combining the results obtained above, we can formulate the main
result:

{\bf Proposition 1:} Every transitive
${\cal G}^{\uparrow}_{+}$-
symplectic manifold is
${\cal G}^{\uparrow}_{+}$-
symplectomorphic with one of the form
$({\cal G}^{\uparrow}_{+}/H,\Omega^{\sigma})$
where $H$ and
$\sigma$
can be:

(1)
$$
H^{1} = \{(R,{\bf 0},\tau,{\bf 0})\vert R \in SO(2),~\tau
\in \R\}
$$
$$
\sigma^{1} = [m,F,S,{\bf 0},{\bf 0}]~~m,S \in \R^{*},~F \in \R
$$

(2)
$$
H^{2} = \{(R(2\pi k/r),{\bf 0},n\gamma,{\bf 0})\vert k=0,...,r-1,
n \in {\bf Z}\}
$$
$$
\sigma^{2} = [m,F,0,{\bf 0},{\bf 0}]~~m\in \R^{*},~F \in \R,~r
\in \N^{*},~\gamma \in \R_{+} \cup \{ 0\}
$$

(3)
$$
H^{3} = \{({\bf 1},v{\bf e}_{1}+\gamma_{1}n{\bf e}_{2},0,
\gamma_{2}m{\bf e}_{1}+a{\bf e}_{2})\vert v,a \in \R,~m,n \in
{\bf Z}\}
$$
$$
\sigma^{3} =[0,0,S,\lambda{\bf e}_{1},k{\bf e}_{2}]~~k \in
\R_{+},~\lambda,S \in \R,~\gamma_{1},\gamma_{2} \in \R_{+} \cup \{0\}
$$

(4)
$$
H^{4} = \left\{\left({\bf 1},v{\bf e}_{1},{Fv\over k},\left( {SFv\over
k^{2}}+{Fv^{2}\over 2k^{2}}+\gamma n\right){\bf e}_{1}+a{\bf e}_{2}\right)
\vert v,a \in \R,~n \in {\bf Z}\right\}
$$
$$
\sigma^{4} =[0,F,S,{\bf 0},k{\bf e}_{2}]~~k \in
\R_{+},~S \in \R,~F \in \R^{*},~\gamma \in \R_{+} \cup \{0\}
$$

(5)
$$
H^{5} = \left\{\left({\bf 1},\left( n\gamma-{S\tau\over k}\right){\bf
e}_{1}+v{\bf e}_{2},\tau,a{\bf e}_{2}\right)\vert \tau,v \in \R,~{\bf
a} \in \R^{2},~n \in {\bf Z}\right\}
$$
$$
\sigma^{5} =[0,0,S,k{\bf e}_{2},{\bf 0}]~~k \in \R_{+},~S \in \R,
{}~\gamma \in \R_{+} \cup \{0\}
$$

(6)
$$
H^{6} = \{(R(2\pi k/r),{\bf v},n\gamma,{\bf a})\vert k=0,...,r-1,~
{\bf v},{\bf a} \in \R^{2},~n \in {\bf Z}\}
$$
$$
\sigma^{6} = [0,0,S,{\bf 0},{\bf 0}]~~S\in \R^{*},~r\in \N^{*},
{}~\gamma \in \R_{+} \cup \{ 0\}
$$

(7)
$$
H^{7} = \{(R(2\pi k/r),{\bf 0},n\gamma,{\bf a})\vert k=0,...,r-1,~
{\bf a} \in \R^{2},~n \in {\bf Z}\}
$$
$$
\sigma^{7} = [0,F,S,{\bf 0},{\bf 0}]~~F,S\in \R^{*},~r\in \N^{*},
{}~\gamma \in \R_{+} \cup \{ 0\}
$$

(8)
$$
H^{8} = \{(R,{\bf 0},\tau,{\bf a})\vert R \in SO(2),~\tau
\in \R,~{\bf a} \in \R^{2}\}
$$
$$
\sigma^{8} = [0,F,0,{\bf 0},{\bf 0}]~~F \in \R^{*}
$$

For distinct couples
$(H,\Omega) \not= (H',\Omega')$
the corresponding manifolds are not
${\cal G}^{\uparrow}_{+}$-
symplectomorphic.

\subsection*{D. Central Extensions of
${\cal G}^{\uparrow}_{+}$}

In principle, the analysis of the transitive symplectic
manifolds for
${\cal G}^{\uparrow}_{+}$
was completed above. However, it is interesting to exhibit such
manifolds as coadjoint orbits. For this we need
$H^{2}({\cal G}^{\uparrow}_{+},\R)$.
We compute this group taking advantage of the knowledge of
$H^{2}({\widetilde{\cal G}^{\uparrow}_{+}},\R)$
which was determined in [2]. For the definitions of
${\widetilde{\cal G}^{\uparrow}_{+}}$
and of the covering map
$\delta:{\widetilde{\cal G}^{\uparrow}_{+}} \rightarrow
{\cal G}^{\uparrow}_{+}$
see also [2].

Let
$c \in Z^{2}({\cal G}^{\uparrow}_{+},\R)$
be arbitrary. We define
$\widetilde{c}:{\widetilde{\cal G}^{\uparrow}_{+}} \times
{\widetilde{\cal G}^{\uparrow}_{+}} \rightarrow \R$
by:
$$
\widetilde{c}(\widetilde{g},\widetilde{g}') \equiv
c(\delta(\widetilde{g}),\delta(\widetilde{g}')).
$$

Then it is elemntary to show that
$\widetilde{c} \in Z^{2}({\widetilde{\cal G}^{\uparrow}_{+}},\R)$.
Applying the result obtained in [2] it follows that
$\widetilde{c}$
is cohomologous with
$m\widetilde{c_{0}}+F\widetilde{c_{1}}+S\widetilde{c_{2}}$
where:
$$
\widetilde{c_{0}} (\widetilde{g},\widetilde{g}') \equiv
{1\over 2} [{\bf a}\cdot R(x){\bf v'}-{\bf v}\cdot R(x){\bf a'}+
\tau' {\bf v}\cdot R(x){\bf v'}].\eqno(4.12)
$$
$$
\widetilde{c_{1}} (\widetilde{g},\widetilde{g}') \equiv
{1\over 2} <{\bf v},R(x){\bf v'}>\eqno(4.13)
$$
$$
\widetilde{c_{2}} (\widetilde{g},\widetilde{g'}) \equiv
\tau x'.\eqno(4.14)
$$

Here
$\widetilde{g} = (x,{\bf v},\tau,{\bf a}),
\widetilde{g'} = (x',{\bf v'},\tau',{\bf a'})$.
Explicitely we have:
$$
c(\delta(\widetilde{g}),\delta(\widetilde{g}')) =
m \widetilde{c_{0}} (\widetilde{g},\widetilde{g}')+
F\widetilde{c_{1}} (\widetilde{g},\widetilde{g}')+
S\widetilde{c_{2}} (\widetilde{g},\widetilde{g}')+
\widetilde{d}(\widetilde{g})-
\widetilde{d}(\widetilde{g}\widetilde{g}')+
\widetilde{d}(\widetilde{g}')\eqno(4.15)
$$
with
$\widetilde{d}:{\widetilde{\cal G}^{\uparrow}_{+}} \rightarrow \R$
a smooth function. By redefining
$\widetilde{d} \rightarrow \widetilde{d'}$
where:
$$
\widetilde{d'}(x,{\bf v},\tau,{\bf a}) =
\widetilde{d}(x,{\bf v},\tau,{\bf a}) -
{x \over 2\pi} \widetilde{d}(2\pi,{\bf 0},0,{\bf 0})
$$
we still have (4.15) but the new
$\widetilde{d}$
also verifies
$$
\widetilde{d}(2\pi,{\bf 0},0,{\bf 0}) = 0.
$$

If we make
$x \rightarrow x+2\pi$
in (4.15) we get more, namely that the function
$\widetilde{d}$
is periodic in $x$ with period
$2\pi$.
Finally, if we make
$x' \rightarrow x'+2\pi$
in (4.15) we get
$S = 0$.
Then it follows from (4.15) that:
$$
c(g,g') = m c_{0}(g,g') + Fc_{1}(g,g') + d(g) - d(gg') + d(g')
$$
where
$$
c_{0} (g,g') \equiv {1\over 2} ({\bf a}\cdot R{\bf v'}-
{\bf v}\cdot R{\bf a'}+ \tau' {\bf v}\cdot R{\bf v'}).\eqno(4.16)
$$
$$
c_{1} (g,g') \equiv {1\over 2} <{\bf v},R{\bf v'}>\eqno(4.17)
$$
with
$g = (R,{\bf v},\tau,{\bf a}), g' = (R',{\bf v'},\tau',{\bf a'})$
and $d$ is uniquely determined by
$d\circ\delta = \widetilde{d}$.

{\bf Proposition 2:}
$H^{2}({\cal G}^{\uparrow}_{+},\R)$
is a two dimensional real space. Explicitely, every $2$-cocycle of
${\cal G}^{\uparrow}_{+}$
is cohomologous with one of the form
$mc_{0} + Fc_{1}$.
\vskip 0.5truecm

According to Subsection 2D we must consider the central
extension
$({\cal G}^{\uparrow}_{+})^{c}$
where
$c = mc_{0} + Fc_{1}$.

It is to be expected that we will obtain only the symplectic
manifolds corresponding to
$S = 0$
from the list included in the statement of Proposition 1.

\subsection*{E. Coadjoint Orbits of
$({\cal G}^{\uparrow}_{+})^{c}$}

We must first compute the function
$\alpha_{g}$
according to (2.17). An elementary computation gives:
$$
\alpha_{R,{\bf v},\tau,{\bf a}}(\alpha,{\bf u},t,{\bf x}) =
$$
$$
m\left( R^{-1}{\bf a}\cdot {\bf u} - R^{-1}{\bf v}\cdot {\bf x} -
\alpha <{\bf a},{\bf v}> - {1 \over 2}t{\bf v}^{2}\right) +
F\left( {1\over 2} \alpha{\bf v}^{2} + <R^{-1}{\bf v},{\bf
u}>\right).\eqno(4.18)
$$

To compute the extended action (2.16) we identify
$(Lie({\cal G}^{\uparrow}_{+})^{c})^{*} \simeq \R +
\R^{2} + \R + \R^{2}$
using the duality form
$$
<(\beta,{\bf G},E,{\bf P}),(\alpha,{\bf u},t,{\bf x})> = -\beta
\alpha - {\bf G}\cdot {\bf u} -Et + {\bf P}\cdot {\bf x}.\eqno(4.19)
$$

Then (2.18) gives for the modified coadjoint action:
$$
Ad^{c}_{R,{\bf v},\tau,{\bf a}}(\beta,{\bf G},E,{\bf P}) =
$$
$$
( \beta+<R{\bf G}+m{\bf a},{\bf v}>-<R{\bf P},{\bf a}>-{1 \over
2}F{\bf v}^{2},R{\bf G}+m{\bf a}-\tau(R{\bf P}+m{\bf v})-FA{\bf v},
$$
$$
E+R{\bf P}\cdot {\bf v}+{1 \over 2}m{\bf v}^{2},R{\bf
P}+m{\bf v}).\eqno(4.20)
$$

It is elementary to compute the orbits of this action. They are:

(a)
$
{\cal O}^{*1}_{m,s,F,{\cal E}} \equiv \left\{(\beta,{\bf G},E,{\bf
P})\vert~~\vert{\bf P} \wedge {\bf G}-(FE-m\beta){\bf e}_{1} \wedge
{\bf e}_{2}\vert=ms,~E-{{\bf P}^{2}\over 2m}={\cal E}\right\}
$

$~m \in\R^{*},~ s \in \R_{+}\cup \{ 0\},~F,{\cal E} \in \R$

(b)
$
{\cal O}^{*2}_{k,\lambda,F} \equiv \{(\beta,{\bf G},E,{\bf
P})\vert~{\bf P} \wedge {\bf G}=(FE-m\beta){\bf e}_{1} \wedge
{\bf e}_{2}, {\bf P}^{2}=k^{2}\}
$

$m=0,~k \in \R_{+},~\lambda,F \in \R$

(c)
$
{\cal O}^{*3}_{E,s,F} \equiv \{(\beta,{\bf G},E,{\bf
0}) \vert~2F\beta+{\bf G}^{2}=s\}
$

$m=0,~s,E,F \in \R$

Now it is easy to match these coadjoint orbits with symplectic
manifolds appearing in the statement of Proposition 1 and
corresponding to
$S = 0$.
We will get maximal manifolds as it is easy to anticipate.
Namely we have:

-
${\cal O}^{*1}_{m,s,F,{\cal E}} $
is for any
$s, {\cal E}$
the maximal manifold of case (2)

-
${\cal O}^{*2}_{k,\lambda,F}$
is the maximal manifold of case (3) and respectively (4) (both for
$S = 0$).

-
${\cal O}^{*3}_{E,s,F}$
is for any
$E,s$
the maximal manifold of case (5)(for
$S = 0$)
and respectively (8).

\subsection*{F. Maximal Symplectic Manifolds for
$S \not= 0$}

- To obtain case (1) we give another realization of case (2).
Namely
$M = \R^{2} \times \R^{2}$
with coordinates
$({\bf q},{\bf p})$,
the symplectic form:
$$
\Omega = \sum_{i=1}^{2} dq_{i} \wedge dp_{i} + {2F\over m^{2}}
dp_{1} \wedge dp_{2}\eqno(4.21)
$$
and the action of
${\cal G}^{\uparrow}_{+}$:
$$
\phi_{R,{\bf v},\tau,{\bf a}}({\bf q},{\bf p}) = \left( R{\bf
q}+{\bf a}-{\tau\over m}(R{\bf p}+m{\bf v}),R{\bf p}+m{\bf
v}\right).\eqno(4.22)
$$

One can see in [7] the corresponding 1+3-dimensional case.
Taking a suggestion from this case we build case (1) as follows.
$M = \R^{2} \times \R^{2} \times S^{1} \times \R$
with coordinates
$({\bf q},{\bf p},{\bf \nu},l)$,
the symplectic form:
$$
\Omega = \sum_{i=1}^{2} dq_{i} \wedge dp_{i} + {2F\over m^{2}}
dp_{1} \wedge dp_{2}+{S\over m} d\varphi \wedge dl\eqno(4.23)
$$
(where
${\bf \nu} = (cos(\varphi),sin(\varphi))$,
and the action
$$
\phi_{R,{\bf v},\tau,{\bf a}}({\bf q},{\bf p},{\bf \nu},l) =
 \left( R{\bf q}+{\bf a}-{\tau\over m}(R{\bf p}+m{\bf v}),
R{\bf p}+m{\bf v},R{\bf \nu},l+m\tau\right).\eqno(4.24)
$$

- In the cases (3)-(5) for
$S \not= 0$,
the trick is to modify a little the extended action (4.20),
namely:
$$
Ad^{c,S}_{R,{\bf v},\tau,{\bf a}}(\beta,{\bf G},E,{\bf P}) =
$$
$$
\left( \beta+<R{\bf G},{\bf v}>-<R{\bf P},{\bf a}>-{1 \over 2}
F{\bf v}^{2}-S\tau,R{\bf G}-\tau R{\bf P}-FA{\bf v},
E+R{\bf P}\cdot {\bf v},R{\bf P}\right).\eqno(4.25)
$$
and to keep the KSK-symplectic form unchanged.

- Case (6) can be realized by:
$M = \R \times S^{1}$
with coordinates
$(T,{\bf \nu})$,
the symplectic form
$$
\Omega = S d\varphi \wedge dT.\eqno(4.26)
$$
and the action
$$
\phi_{R,{\bf v},\tau,{\bf a}}(T,{\bf \nu}) =
(T+\tau,R{\bf \nu}).\eqno(4.27)
$$

-Finaly case (7) is given by
$M = \R \times \R^{2} \times S^{1}$
with coordinates
$(T,{\bf V},{\bf \nu})$,
the symplectic form
$$
\Omega = F dV_{1} \wedge dV_{2}+S d\varphi \wedge dT+k(-sin(\varphi)
dV_{1}+cos(\varphi) dV_{2}) \wedge dT\eqno(4.28)
$$
and the action
$$
\phi_{R,{\bf v},\tau,{\bf a}}(T,{\bf V},{\bf \nu}) =
(T+\tau,R{\bf V}+{\bf v},R{\bf \nu}).\eqno(4.29)
$$

Of course in (4.26) and (4.28)
${\bf \nu} = (cos(\varphi),sin(\varphi))$
as in case (1).

\subsection*{G. Factorized Symplectic Manifolds}

In all cases except (1) and (8) we have, beside the maximal
manifolds exhibited above, some families of factorized
manifolds. We indicate briefly the results in these remaining
cases.

(2) If
$\gamma = 0$
and
$r=2,3...$
then we modify (4.24) as follows;
$$
\phi_{R,{\bf v},\tau,{\bf a}}({\bf q},{\bf p},{\bf \nu},l) =
 \left( R{\bf q}+{\bf a}-{\tau\over m}(R{\bf p}+m{\bf v}),
R{\bf p}+m{\bf v},R^{r}{\bf \nu},l+m\tau\right).\eqno(4.30)
$$

To obtain the cases with
$\gamma \in \R_{+}$
one factorizes the previous cases to the following action of
{\bf Z}:
$$
n\cdot ({\bf q},{\bf p},{\bf \nu},l) =
({\bf q},{\bf p},{\bf \nu},l+n\gamma m).\eqno(4.31)
$$

(3) One factorizes the maximal manifolds to an action of
${\bf Z} \times {\bf Z}$
namely:
$$
(n,m)\cdot (\beta,{\bf G},E,{\bf P}) =
(\beta+mk\gamma_{2},{\bf G},E+nk\gamma_{1},{\bf P}).\eqno(4.32)
$$

(4)-(5) The maximal manifold is factorized to the following
action of
{\bf Z}:
$$
n\cdot (\beta,{\bf G},E,{\bf P}) =
(\beta+nk\gamma,{\bf G},E,{\bf P}).\eqno(4.33)
$$

(6) If
$\gamma = 0$
and
$r=2,3,...$
we proceed as at (2) above, namely we modify (4.27) to:
$$
\phi_{R,{\bf v},\tau,{\bf a}}(T,{\bf \nu}) =
(T+\tau,R^{r}{\bf \nu}).\eqno(4.34)
$$

The case
$\gamma \in \R_{+}$
is obtained factorizing the previous cases to the following
action of {\bf Z}:
$$
n\cdot (T,{\bf \nu}) = (T+n\gamma,{\bf \nu}).\eqno(4.35)
$$

(7) Similarly with (6): the action on the variable {\bf V} is
not changed.

{\bf Remark 5:} The notion of localisability can be investigated
as in [9]. It is manifest from the first part of Subsection F
that cases (1) and (2) (i.e. non-zero mass systems) are
localisable on
$\R^{2}$.
One also finds out that the cases (5)-(7) are localisable on
$S^{1}$.

{\bf Remark 6:} The existence of a momentum map is more subtle
than in the case of the Poincar\'e group. Namely for
$S \not= 0$
such a map does not exists and for
$S = 0$
a momentum map for the maximal manifolds is the identity map in
the coadjoint representation from Subsection 4E. If
$S = 0$
but we are dealing with a factorized manifold, again the
momentum map does not exists.

\end{document}